# Nitrogen Isotopic Composition and Density of the Archean Atmosphere

Bernard Marty[1*], Laurent Zimmermann[1], Magali Pujol[1†], Ray Burgess[2], Pascal Philippot[3]

[1] CRPG-CNRS, Université de Lorraine, 15 rue Notre Dame des Pauvres, 54501 Vandoeuvre-lès-Nancy Cedex, France

[2] School of Earth, Atmospheric and Environmental Sciences, University of Manchester, Oxford Road, Manchester, M13 9PL, United Kingdom.

[3] Institut de Physique du Globe de Paris, Sorbonne-Paris Cité, Université Paris Diderot, CNRS, 1 rue Jussieu 75238 Paris Cedex 5, France.

\* Corresponding author: bmarty@crpg.cnrs-nancy.fr
† Present address : Fluid and Organic Geochemistry, TOTAL Exploration & Production, Avenue Larribau 64000 PAU, France

Title : 67 characters

Abstract : 95 words

Main text : 1903 words

30 references

2 Figures

Revision, September 5th, 2013

**Understanding the atmosphere's composition during the Archean eon is a fundamental issue to unravel ancient environmental conditions. We show from the analysis of nitrogen and argon isotopes in fluid inclusions trapped in 3.0-3.5 Ga hydrothermal quartz that the $P_{N2}$ of the Archean atmosphere was lower than 1.1 bar, possibly as low as 0.5 bar, and had a nitrogen isotopic composition comparable to the present-day one. These results imply that dinitrogen did not play a significant role in the thermal budget of the ancient Earth and that the Archean $P_{CO2}$ was probably lower than 0.7 bar.**

Nitrogen is a key element of planetary atmospheres that constitutes a sensitive tracer of exchanges of volatile elements between planetary interiors and the outer space *(1-3)*. On Earth, the amount of nitrogen at the Earth's surface might have varied significantly with time, as a result of exchange with the deep Earth *(1, 3, 4)* or of loss to space *(2, 6)*. In contrast, the amount of non-radiogenic (not produced by nuclear reactions) noble gas isotopes in the terrestrial atmosphere is unlikely to have varied significantly since the Earth's formation *(7)*. Calibrating nitrogen against a non-radiogenic noble isotope such as $^{36}Ar$ in ancient samples provides a way to estimate the variations of the atmospheric $N_2$ partial pressure in the past.

We estimate here the $N_2$ partial pressure in the ancient terrestrial atmosphere from the analysis of the $N_2/^{36}Ar$ ratios in well characterized fluid inclusions trapped in hydrothermal quartz from the Archean 3.49 Ga Dresser *(8-11)* and 3.46 Ga Apex Basalt *(12)* formations, Pilbara Craton (Western Australia). Previous geochemical studies of fluid inclusions from this area *(10, 12-16)* and from other geological settings *(18)* have shown the presence of a paleo-atmospheric component, likely incorporated as atmospheric gases dissolved in surface waters. The abundances of atmospheric gases in water are a direct function of their respective partial pressures and of salinity and temperature of water *(19)*. For the present-day atmospheric partial pressures of $N_2$ (0.7906 bar) and $^{36}Ar$ (3.20 x $10^{-5}$ bar), the $N_2/^{36}Ar$ ratio of air-saturated water (ASW) ranges from 1.02 x $10^4$ to 1.31 x $10^4$, for water temperatures between 2°C (the present-day average deep-sea temperature) and 70°C (a proposed model temperature for the Archean oceans *(20)*), and salinities between 0 and 16 wt % eq. NaCl (encompassing the range of salinities observed in Archean the Dresser and Apex fluid inclusions *(10, 12)*). Nishizawa et al. *(13)* have shown that, in fluid inclusions trapped in silica dykes and quartz veins from the Dresser Formation, $N_2$ is the main nitrogen species, far more abundant than the other identified species ($NH_4^+$). These authors proposed an upper limit of 3.3 times the modern value for the Archean $N_2/^{36}Ar$ ratio.

Fluid inclusions trapped in three different Archean hydrothermal quartz samples were analyzed *(10-12, 15, 16)*. The choice of the samples was driven by the goal of targeting pristine fluids of marine and/or meteoric origin. Ideally, the best samples to be investigated would be sedimentary rocks. However, precipitated minerals, e.g. carbonates, sulfates, halides, are relatively weak mineral phases compared to quartz, seldom preserving primary fluid inclusions. Silicified sediments were another option but the grain-size is generally too small to preserve fluid inclusions large enough to analyze (several μm). For this reason we chose to investigate quartz-bearing amygdules preserved in komatiitic basalt flow (PI-06) at the base of the Dresser Formation and recovered through drilling, and quartz-bearing pods filling retreat voids in exposed pillow basalts (PI-02-39) at the top of the Dresser Formation and within the Apex Formation in the Warranoowa syncline, (PB-02-122). Considering the high silica content of Archean seawater and hydrothermal fluids, such voids likely were filled with quartz immediately upon cooling. Recognition that the intrapillow pods form well defined ovoid shapes isolated in the core of the pillow indicates that fluid circulation processes driving mineral precipitation should have occurred through a porous medium shortly after basalt deposition. In order to maximize sampling of primary fluids of surficial origin (seawater, meteoric) and to minimize the imprint of hydrothermal fluids, all samples were collected in undeformed rocks located in close proximity to the overlying marine/lacustrine sediments. This, together with the shallow water character of both the Dresser and Apex settings, indicates that the samples investigated contain fluids of marine, and/or lacustrine origin(s), in addition to hydrothermal fluids.

PI-06 quartz sample is from a drill core (Pilbara Drilling Project 2) in the Dresser formation and was selected in the depth interval 102-110 m. The quartz fills 2–10 mm vesicles in komatiitic basalt at the base of the Dresser Formation, and contains 2-10 μm-sized fluid inclusions. The formation of quartz in amygdules in komatiitic basalt must have occurred early in the lava post-emplacement history since neither deformation nor pressure tracks were observed *(15)*. Fluids trapped in the inclusions have been interpreted as being a mixture of Archean surface water and hydrothermal fluid *(15, 17)*. Ar-Ar analysis of trapped fluids indicates that they are ≥3.0±0.2 Ga-old *(15)* and that they contain inherited $^{40}Ar$, presumably from an hydrothermal end-member.

Samples PI-02-39 and PB-02-122 are from isolated quartz-carbonate aggregates forming pods hosted in pillow basalts now exposed at the surface. These pods resemble typical mineralisation structures associated with passive hydrothermal circulation of water through shallow crust. Intra-pillow quartz crystals, which were selected for analysis, contain abundant, 1-25 μm, two phase (liquid and <5% $CO_2$ vapor) aqueous inclusions *(10)*. Fluid Inclusions are randomly distributed throughout the host quartz, which argues for a primary origin. The absence of crosscutting veins, metamorphic overprint, and deformation features affecting pillow basalts and associated pods indicates negligible fluid remobilization and circulation after deposition and crystallization. Foriel et al. *(10)* demonstrated that several fluids were trapped in PI-02-39, such as an evolved Archean water component and several (Ba, Fe-rich) fluid components. A previous noble gas (Ar and Xe) study of sample PI-02-39 *(16)* has shown that the quartz has preserved since the meso-Archean era a paleo-atmospheric component having a $^{40}Ar/^{36}Ar$ ratio of 143±24, mixed with an hydrothermal fluid component rich in Cl, K and inherited $^{40}Ar$. The Ar-Ar analysis of PI-02-39 indicates that fluids are >2.7 Ga-old, probably contemporaneous to the deposition of the Dresser formation *(16)*. Although not investigated in details for its fluid inclusion composition, sample PB-02-122 shows the same type of fluid inclusion distribution and texture as sample PI-02-39. Further sample characteristics field locations and photos are given in the Supplementary Online *(17)*.

Several aliquots of each sample were crushed under vacuum using different numbers of crushing steps and the extracted gases were sequentially analyzed for N and Ar isotopic abundances by static mass spectrometry (Table S1). For all runs, the nitrogen and argon abundances correlate, showing that it is unlikely that much nitrogen in the fluid inclusions has been consumed by post-entrapment chemical reactions. Prior to analysis, two aliquots of sample PI-02-39 were neutron-irradiated, in order that the extended Ar-Ar technique *(21)* could be used to determine K and Cl in the same extractions as N and Ar. The combined analyses of chlorine, potassium, argon and nitrogen isotopes confirm previous studies *(10, 13-16)* that trapped fluids are mixtures of a low salinity, low N, and low radiogenic $^{40}Ar$ end-member, with several hydrothermal components rich in N, Cl, K and radiogenic $^{40}Ar$.

For all samples, step-crushing data define well resolved mixing correlations in a $^{40}Ar/^{36}Ar$ vs. $N_2/^{36}Ar$ diagram that are consistent with mixing between several hydrothermal fluids and a low $^{40}Ar/^{36}Ar$, $N_2/^{36}Ar$ end-member (Fig. 1). The slopes of the correlations differ widely between samples, indicating variable enrichment of $^{40}$Ar in the different trapped hydrothermal end-members. The lowest measured N$_2$/$^{36}$Ar ratios in step-crushing runs of PI-02-39 aliquots range from 0.72±0.03 x 10$^4$ (last crushing step of PI-02-39-3, Table S1 & S3) to 1.43±0.02 x 10$^4$ (first crushing step of PI-02-39-4). A higher value of 3.13±0.02 x 10$^4$ was obtained for aliquot PI-02-39-2, however we suspect the contribution of hydrothermal nitrogen in this case *(17)*. We consider a N$_2$/$^{36}$Ar range of 0.7-1.4 x 10$^4$ as representative of the low salinity end-member of sample PI-02-39. Sample PI-06 presents much lower $^{40}$Ar/$^{36}$Ar ratios in the range 386-1,014 than those of sample PI-02-39, in agreement with its less evolved character of its trapped fluids *(15, 17)*. Despite this difference, its lowest N$_2$/$^{36}$Ar ratios (0.85-1.50 x 10$^4$, Table S3) are in the same range as those of sample PI-02.

The convergence of data points towards a common end-member (Fig. 1) constrains the possible range of N$_2$/$^{36}$Ar ratios for the presumed Archean ASW value, which is consistent with the modern ASW value. Based on the results summarized in Table S3, we propose that the Archean ASW N$_2$/Ar ratio was ≤ 0.7-1.4 x 10$^4$, comparable to that of modern ASW (1.0-1.3 x 10$^4$). Assuming a constant concentration of atmospheric $^{36}$Ar since 3.5 Ga *(7)* implies that the Archean P$_{N2}$ was 1.1 bar (scaled to the modern P$_{N2}$ of 0.79 bar), and possibly as low as 0.5 bar.

Variations of the nitrogen isotopic composition (expressed as δ$^{15}$N relative to the modern atmospheric value) are also consistent with mixing between a crustal/sedimentary end-member (δ$^{15}$N within 3-8‰ for metagabbros *(4)*, and 5-15‰ for sediments (e.g., *4, 5, 14, 22*), and an Archean atmospheric δ$^{15}$N value within ~2-3‰ of the present-day value (Fig. 2). A modern-like N isotope ratio in the Archean is in agreement with the conclusions of a near-constant atmospheric δ$^{15}$N through time from the analysis of ancient cherts *(14, 22)*, although others *(23)* have proposed drastic $^{15}$N enrichments of atmospheric N$_2$ during the Archean eon. The near-constancy of the atmospheric δ$^{15}$N value and of the P$_{N2}$ since 3.0-3.5 Ga is consistent with the presence of a significant terrestrial magnetic field in the Archean. In the absence of such magnetic shielding, atmospheric dinitrogen would have interacted with charged particles from the solar wind, resulting in non-thermal loss to space of this element and N isotope fractionation *(3)*, as in the atmospheres of Mars *(24)* and Titan *(25)*. In order to shield efficiently the amosphere against N$_2$ loss, a magnetic field of at least 50 % the present-day intensity is required 3.5 Ga ago *(3)*, in agreement with paleomagnetic data from 3.2 Ga single silicate crystals *(26)*.

An Archean atmospheric nitrogen partial pressure ≤ 1.1 bar together with a N isotopic composition similar to that of the present-day atmosphere has important implications for the thermal conditions of the Earth's surface. The energy delivered by the ancient Sun might have been as little as 70 % of what is today, requiring other sources of energy or, more probably, a higher greenhouse effect than today *(27)*. The presence of greenhouse gases such as $NH_3$ and $CH_4$ has been proposed (e.g., *(28)* and refs therein) but their stability in the ancient atmosphere has been questioned. Others have postulated the occurrence of higher pressures of $CO_2$ in the past, although the geological record of Archean sedimentary rocks suggests that the $P_{CO_2}$ could have been only a few times the present-day value *(28)*. Two studies *(1, 2)* have proposed that a $P_{N2}$ 2-3 times the present-day one (with the presence of other gaseous species like $H_2$ *(2)*) was sufficient to maintain a clement surface temperature with partial pressures of greenhouse gases consistent with the geological record. The present results are not consistent with these proposals and can be used to set an upper limit of the maximum Archean $P_{CO_2}$ pressure. A recent study *(29)* based on fossil imprints of rain droplets proposed an atmospheric pressure < 2 bar at 2.7 Ga, probably less than an upper limit in the range 0.5-1.14 bar, which is similar to our estimate of the $P_{N2}$ in the Archean atmosphere (0.5-1.1 bar). Although the ages of our Archean samples may differ, this comparison suggests by difference between the two estimates that the partial pressure of $CO_2$ was not more than ~0.7 bar and may have been far less. This is qualitatively consistent with conditions necessary to maintain a temperature of ~15 °C at the Earth's surface with a mixture of $CO_2$ and other greenhouse gases *(27, 28)*.

.

30. This study was supported by ANR Grant eLife-2 to PP and BM, and by the European Research Council under the European Community's Seventh Framework Programme (FP7/2007-2013 grant agreement no. 267255 to BM). PP acknowledges the Institut de Physique du Globe de Paris, the Institut des Sciences de l'Univers, and the Geological Survey of Western Australia for supporting the Pilbara Drilling Project. We thank six anonymous reviewers and David Catling for constructive comments. This is CRPG contribution #2248 and IPGP contribution #3424.

**Supplementary Materials**
www.sciencemag.org
Materials and Methods
Figs. S1 to S4
Tables S1 to S3
References *(31-38)*

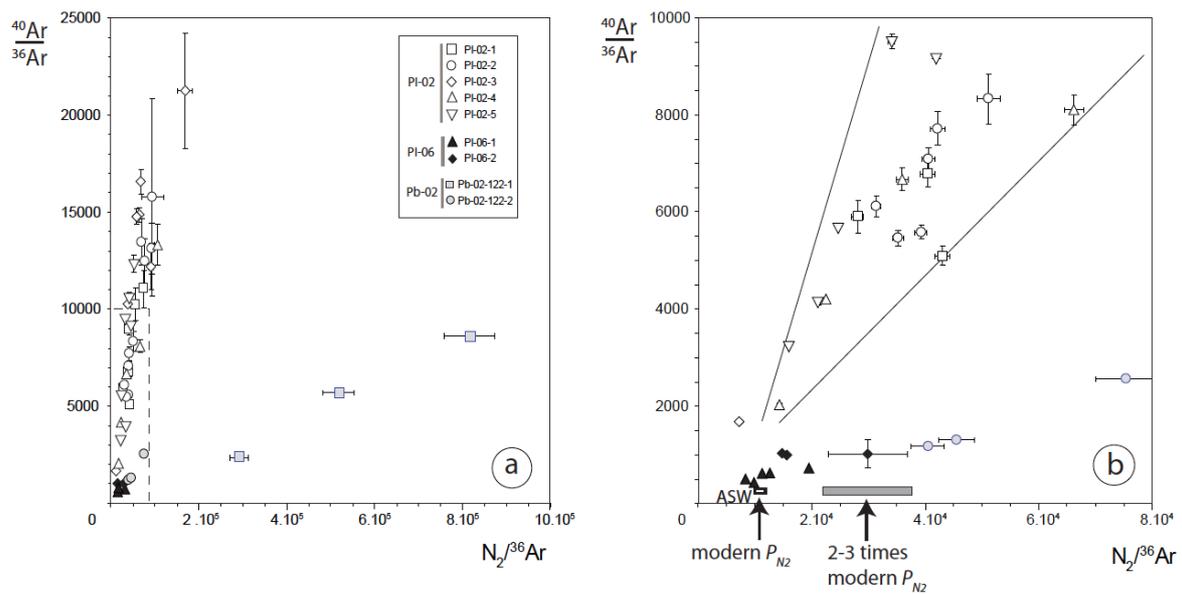

**Figure 1 :** N-Ar isotope variations for inclusion fluids trapped in Archean quartz. Open, black and grey symbols are duplicate step-crushing runs for samples PI-02-39-39, PI-06, and PB-02-122, respectively. Fig. 2b is an enlargement of the dotted zone of Fig. 2a. Lines in 2b represent upper and lower limits for sample PI-02-39 data points which converge towards ASW. This sample contains a mixture of several hydrothermal fluids having different N/$^{36}$Ar compositions with a Cl-poor water component. ASW : air-saturated water with N$_2$/$^{36}$Ar values corresponding to modern atmospheric partial pressures of nitrogen and argon (Table S3), and with $^{40}$Ar/$^{36}$Ar ratios < 298 (the modern value). The corresponding modern partial pressure of atmospheric N$_2$ is also indicated. A 2-3 times higher N$_2$ pressure, as suggested by *(1, 2)*, would correspond to the gray bar on the right hand side of ASW. Such a higher value is not supported by the convergence of data point correlations.

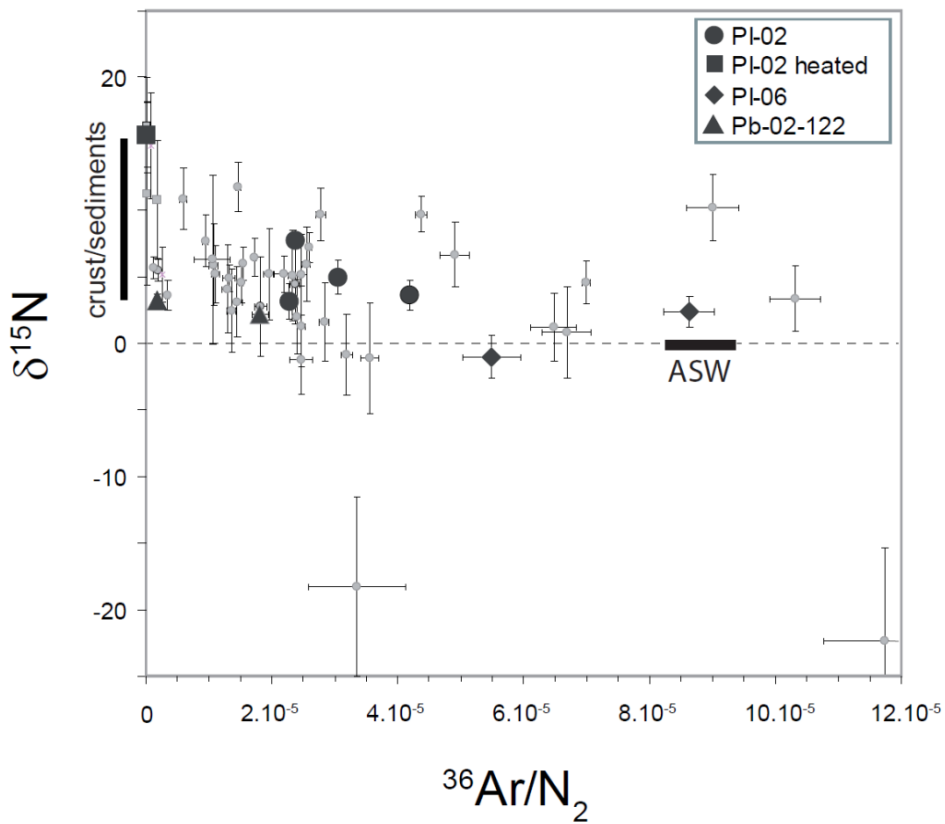

**Figure 2 :** Nitrogen isotopic composition ($\delta^{15}N$ is the deviation in part per mil from the modern atmospheric $^{15}N/^{14}N$ ratio of $3.6765 \times 10^{-3}$) versus the $^{36}Ar/N_2$ ratio for all extraction steps (small gray symbols) and for the total extracted gases of each samples (large black symbols). In this format, mixings will yield straight lines. All extractions were done by crushing except a heating run for PI-02-39 (gray and black squares). ASW is the modern air-saturated water composition ($^{36}Ar/N_2$ computed with data given in Table S3, $\delta^{15}N$ = 0 ‰). The range of crustal and sedimentary values is also indicated *(4, 5, 14, 22)*. Observed ratios in Archean fluid inclusions are consistent with mixing between typical crustal/sedimentary values from the hydrothermal end-members and an ASW component having a N isotopic composition comparable to the modern one within ~2-3‰.